# A Platform for Implementing Secure Wireless Ad Hoc Networks

Gyula FARKAS[1], Béla GENGE[2], Piroska HALLER[3]

Department of Electrical Engineering, Faculty of Engineering,
"Petru Maior" University, Tg. Mureş, e-mail: [1]farkas_gyula@yahoo.com,
[2]bgenge@engineering.upm.ro, [3]phaller@engineering.upm.ro

**Abstract:** We propose a new platform for implementing secure wireless ad hoc networks. Our proposal is based on a modular architecture, with the software stack constructed directly on the Ethernet layer. Within our platform we use a new security protocol that we designed to ensure mutual authentication between nodes and a secure key exchange. The correctness of the proposed security protocol is ensured by Guttman's authentication tests.

## 1. Introduction

Unlike traditional wireless networks, wireless ad hoc networks do not rely on any fixed infrastructures. Nodes communicate with other nodes that are within their radio range and by using routed messages to communicate with nodes that are not directly reachable. This flexibility ensures that wireless networks can be set up in an ad hoc manner where there is no infrastructure available.

Ensuring a secure communication that also provides security properties such as authentication, secrecy, integrity and non-repudiation involves using security protocols [1]. There have been several security protocols and solutions proposed in the literature to ensure security goals [2, 3, 4]. However, these solutions are mainly focused on designing new security protocols and secure routing protocols and do not focus on providing a modular platform where security protocol implementations can be easily replaced with other implementations.

In this paper we propose a solution for the authentication problem of nodes communicating directly and a solution to the secrecy, integrity and non-repudiation of messages exchanged between nodes. Instead of focusing on the security of routing protocols, we provide a platform that ensures security of directly linked nodes with a modular architecture that can be used to implement secure discovery, secure routing, and other applications for wireless ad hoc networks.

In order to ensure mutual authentication and a secure key exchange between directly communicating nodes, we design a new protocol, based on Guttman's authentication tests [5]. This protocol can be replaced with other implementations in order to satisfy the security requirements of applications. The platform requires a Linux OS, the OpenSSL [6] security library and a wireless network adapter for execution.

The paper is structured as follows.

## 2. Platform architecture

The software stack of the proposed platform is shown in *Fig. 1*. Our platform uses the *Ethernet* layer as a base package transfer layer. The advantages of using this layer are multiple. First of al, building ad hoc networks over well known network or transport layers is not possible because of the dynamic of ad hoc networks where nodes can change their location and there is no fixed infrastructure to provide IP addresses or to route packages based on IP addresses.

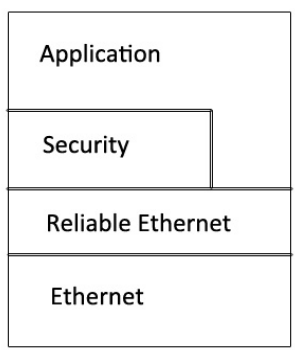

*Figure 1:* Platform software stack.

As another advantage, the overhead that would be generated by IP/TCP/UDP protocols is completely removed. However, this means that we must now deal with package loss/fragmenting. For this purpose we introduce a new layer on top of the simple Ethernet layer: *Reliable Ethernet*.

The *Security* layer is the one that provides an implementation of security goals for the platform. It ensures that nodes are properly authenticated, it ensures secrecy and integrity of data.

Based on the security layer, we can implement several applications on the *Application* layer. Applications can also bypass the security layer if security is not a requirement, or if it is handled at the application layer.

The block diagram of the proposed platform can be seen in *Fig. 2*. The platform consists of several modules. The "Connection Management" module is the core of the platform, it is responsible for creating message headers and maintaining connections. The "Reliable Network" is the implementation of the *Reliable Ethernet* layer. The "Container" implements the management of *devices* representing state machines corresponding to each node that has been discovered.

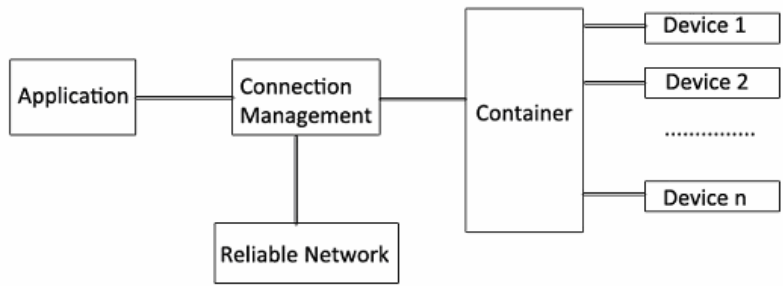

*Figure 2:* Internal block diagram.

## 3. Security considerations

The authentication can be based on password or certificate. We can distinguish three cases, where participants *A* and *B* both have certificates; only *A* has a certificate; no participant has a certificate. In this paper we present the protocol we used for mutual authentication where both parties are in the possession of certificates obtained from a Certificate Authority. The other cases are derived from this protocol, leading to the authentication of only one party, pre-defined key-based authentications and no authentication.

Guttman and Fabrega proposed a method for designing authentication protocols [5], based on authentication tests. In our proposed platform we used this method to design a new mutual authentication protocol that also provides a secure session key exchange. Session keys are used later to provide secrecy of data through symmetric block ciphers and integrity of data through message authentication codes.

The designed security protocol is shown in *Fig. 3*. The initiator of the protocol is *A*, who generated a random number $N_1$ and sends it to *B* along with his certificate $C_A$. *B* generates a random number $N_2$ and a session key *K* which is encrypted with its private key. *B* creates the hash $\{N_1, K, \{K\}_{Pk(A)}\}_h$ and signs it with his private key. Then, *B* sends the message $N_2, K, \{K\}_{Pk(A)}, C_B, \{ \{N_1, K, \{K\}_{Pk(A)}\}_h\}_{Pvk(B)}$. Finally *A* will confirm that he is running the same session by generating and sending to *B* the hash $\{N_1, N_2, K, \{K\}_{Pk(A)}\}$ signed with his private key.

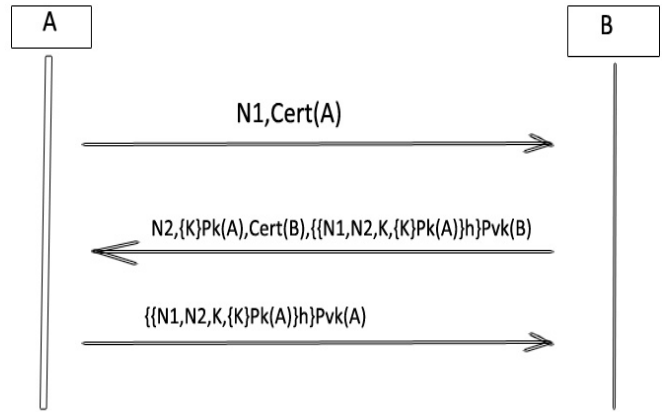

*Figure 3:* Proposed mutual authentication protocol.

## 4. Conclusion

We proposed a new platform for implementing wireless ad hoc networks. The platform has been designed for Linux OS, with a compiled OpenSSL library where a wireless network adapter is available. The proposed platform can be used to implement several application types such as: secure routing, secure data transfer, secure node discovery. Within the platform we use a new security protocol that we designed based on Gutman's authentication tests. The protocol ensures mutual node authentication and secure session key exchange.